\magnification1200
\baselineskip = 16pt
\parskip = 10pt

\font\BBig=cmr10 scaled\magstep2
\font\goth=eufm10 


\def\math#1{\mathop{\rm #1}\nolimits}

\def\Diff{\math{Diff}}
\def\Vect{\math{Vect}}
\def\cor{\mathop{\hbox{\goth{cor}}}}
\def\mil{\mathop{\hbox{\goth{mil}}}}
\def\gal{\mathop{\hbox{\goth{gal}}}}
\def\so{\mathop{\hbox{\goth{so}}}}
\def\se{\mathop{\hbox{\goth{se}}}}
\def\Fun{C^\infty}

\def\parag{\hfil\break}

\def\GammaU{{}^{U}\!\Gamma}
\def\gammaU{{}^{U}\!\gamma}

\def\smallover#1/#2{
\hbox{$\textstyle{#1\over#2}$}
} 

\def\half{{\smallover 1/2}}

\def\semidirectproduct{
{\ooalign
{\hfil\raise.07ex\hbox{s}\hfil\crcr\mathhexbox20D}}
} 

\def\bR{{\bf R}}

\def\cA{{\cal A}}
\def\cF{{\cal F}}

\def\X{X}

\def\ccr{\cr\noalign{\smallskip}}


\centerline{
{\BBig On Galilean Isometries}
}

\bigskip
\bigskip
\centerline{
Christian DUVAL\footnote{*}{CPT-CNRS and
Universit\'e d'Aix-Marseille II}
}

\bigskip
\centerline{Centre de Physique Th\'eorique CNRS, Case 907}
\centerline{F--13288 MARSEILLE Cedex 9, FRANCE}
\centerline{(Christian.Duval@cpt.univ-mrs.fr)}

\bigskip



{\sevenrm\it
We introduce three nested Lie algebras of infinitesimal
`isometries' of a Galilei space\-time structure which play the
r\^ole of the algebra of Killing vector fields of a
relativistic Lorentz space\-time. Non trivial extensions of
these Lie algebras arise naturally from the consideration of
Newton-Cartan-Bargmann automorphisms.
}

\bigskip


\parag
{\bf 0. Introduction}

Quite recently, Carter and Khalatnikov [{\bf{CK}}] have pointed out that a
geometric four\-dimensional formulation of the non relativistic Landau theory of
perfect super\-fluid dynamics should involve not only Galilei covariance but
also, more significantly as far as gravitational effects are concerned,
covariance under a larger sym\-metry group which they call the {\it Milne group}
after Milne's pioneering work in Newtonian cosmology~[{\bf{Mi}}].

The purpose of this note is to show that the degeneracy of the
Galilei `metric' [{\bf{Ca}},{\bf{Tra}},{\bf{K\"u}}] allows for a certain
flexibility in the definition of space\-time `isometries'. More precisely,
{\it three} different nested Lie algebras of space\-time vector fields
naturally arise as candidates for Newton-Cartan symmetry algebras, one of
them being the newly highlighted Milne algebra. This is reviewed in
section~1 in a simple algebraic way.

Section 2 is devoted to a detailed study of the various extensions of
these isometries in the framework of Newton-Cartan-Bargmann structures
associated with a Newtonian principle of general covariance that goes back to
Cartan. These nested Lie algebras actually embody the {\it Bargmann} algebra
which generates the fundamental symmetry group of massive, either classical or
quantum, non relativistic isolated systems.

\goodbreak
\parag
{\bf 1. Coriolis, Galilei and Milne algebras}

Let us recall that a {\it Newton-Cartan} (NC) {\it structure} [{\bf K\"u}] for
space\-time, $(M,\gamma,\theta,\Gamma)$, consists of a smooth manifold $M$ of
dimension $n+1$, a $2$-contravariant semi-positive sym\-metric tensor field
$\gamma=\gamma^{ab}\,\partial_a\otimes\partial_b$
($a,b=0,1,\ldots,n$) whose kernel is spanned by the {\it time} $1$-form
$\theta=\theta_a\,dx^a$; also $\Gamma$ is a torsion-free linear connection
compatible with $\gamma$ and $\theta$. Now such a connection is far from being
uniquely determined by the {\it Galilei structure} $(M,\gamma,\theta)$,
therefore {\it Newtonian connections} are furthermore subject to the nontrivial
sym\-metry of the curvature $R_{a\ c}^{\ b\ d}=R_{c\ a}^{\ d\ b}$ 
(where $R_{a\ c}^{\ b\ d}\equiv\gamma^{bk}\,R_{akc}^{\ \ \ d}$) which may be
thought of as part of the gravitational field equations~[{\bf{DK}}].

The {\it standard} example of a NC structure is given by
$M\subseteq\bR\times\bR^n$ together with
$\gamma=\sum_{A=1}^{n}{\partial_A\otimes\partial_A}$
and $\theta=dx^0$;
the nonzero components of the connection
$\Gamma_{00}^{A}=\partial_A\phi$ ($A=1,\ldots,n$)
accomodate the Newtonian scalar potential $\phi$.

The {\it flat} NC structure corresponds to the trivial case
$\Gamma_{ab}^{\ \ c}\equiv 0$.

\goodbreak
\parag
i) {\sl Coriolis algebra}

The vector fields $\X=\X^a\,\partial_a$ on $M$
that satisfy
$$
L_\X\gamma^{ab} = 0,
\qquad
L_\X\theta_a = 0
\leqno(1)
$$
form an infinite-dimensional Lie algebra called the
{\it Coriolis} algebra $\cor(M,\gamma,\theta)$.
Notably enough, these vector fields do not Lie-transport the
Newtonian connection; nevertheless, a somewhat tedious
calculation, using the above mentioned prescribed symmetry
of the curvature, shows that
$$
L_\X\Gamma^{abc}=0
\leqno(1')
$$
where
$
\Gamma^{abc}
\equiv
\gamma^{ak}\gamma^{b\ell}\Gamma_{k\ell}^{\ \ c}
$.

In the standard case one finds 
$$
\X^A=\omega(t)^A_B\,x^B + \varrho(t)^A,
\qquad
\qquad
\X^0=\tau
\leqno(2)
$$ 
where
$\omega$ (resp. $\varrho$) is some $\so(n)$ (resp. $\bR^n$) valued function of
time $t=x^0$ and $\tau\in\bR$ an infinitesimal time translation.
Coriolis vector fields generate the so called `accelerated frames'
transformations.

\goodbreak
\parag
ii) {\sl Galilei algebra}

The affine Coriolis vector fields $\X$, {\it viz}
$$
L_\X\gamma^{ab} = 0,
\qquad
\qquad
L_\X\theta_a = 0,
\qquad
\qquad
L_\X\Gamma_{ab}^{\ \ c} = 0
\leqno(3)
$$
form the {\it Galilei} Lie algebra
$\gal(M,\gamma,\theta,\Gamma)$ of our Newton-Cartan
structure.
This algebra has maximal dimension $(n+1)(n+2)/2$.

In the flat case we obtain
$$
\X^A=\omega^A_B\,x^B + \beta^A\,t + \sigma^A,
\qquad
\qquad
\X^0=\tau
\leqno(4)
$$
with $(\omega,\beta)\in\se(n)$ and $(\sigma,\tau)\in\bR^{n+1}$.

This general definition, originally due to Trautman [{\bf{Tra}}],
provides a clearcut geometrical status for the fundamental symmetries of
Galilean classical mechanics and field theory.

\goodbreak
\parag
iii) {\sl Milne algebra}

Interestingly, there exists a less familiar intermediate algebra, namely the
infinite-dimensional Lie algebra of those vector fields $\X$ such that
$$
L_\X\gamma^{ab} = 0,
\qquad
\qquad
L_\X\theta_a = 0,
\qquad
\qquad
L_\X\Gamma_a^{\ bc} = 0
\leqno(5)
$$
where $\Gamma_a^{\ bc}\equiv\gamma^{bk}\Gamma_{ak}^{\ \ c}$.
We will call it the {\it Milne} algebra
$\mil(M,\gamma,\theta,\Gamma)$.

In the standard case (which encompasses Newton-Milne cosmology
[{\bf{Mi}}]) we get
$$
\X^A=\omega^A_B\,x^B + \varrho(t)^A,
\qquad
\qquad
\X^0=\tau
\leqno(6)
$$
with the same notation as before.

These vector fields constitute a Lie algebra corresponding to the infinitesimal
Milne transformations introduced in [{\bf{CK}}] which, indeed, admit the
intrinsic definition given by Eqs~(5) for a general Newton-Cartan structure
$(M,\gamma,\theta,\Gamma)$.

\goodbreak
\parag
{\bf 2. Extending the Coriolis, Milne and Galilei algebras}

It has been established [{\bf{K\"u}}] that any newtonian
connection can be affinely decomposed according to
$
\Gamma_{ab}^{\ \ c}
=
\GammaU_{ab}^{\ \ c}+\theta_{(a}F_{b)k}\gamma^{kc}
$
where [{\bf{Tr\"u}}]
$$
{\GammaU}_{ab}^{\ \ c}
=
\gamma^{ck}\Big(
\partial_{(a}\gammaU_{b)k} - \half\partial_{k}\gammaU_{ab}
\Big)
+
\partial_{(a}\theta_{b)}\,U^c 
\leqno(7)
$$
is the unique NC connection for which the unit space\-time vector field $U$
(i.e. $\theta_aU^a=1$) is geodesic and curlfree, $F$ being an otherwise 
arbitrary closed $2$-form, locally
$$
F_{ab}=2\,\partial_{[a}A_{b]}.
\leqno(8)
$$
Here $\gammaU$ is uniquely determined by
$\gammaU_{ak}\gamma^{kb}=\delta_a^b -U^b\theta_a$ and
$\gammaU_{ak}U^k=0$.

As an illustration, standard NC structure corresponds to the gauge choice
$U=\partial/\partial{x}^0$ and $A=-\phi\,\theta$.

So, NC strutures $(M,\gamma,\theta,\Gamma)$ are best represented by what we
call {\it Newton-Cartan-Bargmann} (NCB) structures
$(M,\gamma,\theta,U,A)$ via the previous formul\ae.
Now, experience suggests to think of a NCB structure
as the sextuple $(M,\gamma,\theta,U,V,\phi)$ where
$$
V^a=U^a-\gamma^{ak} A_k
\leqno(9)
$$
is an observer in rotation with respect to the `ether' $U$
[{\bf{CK}}], while
$$
\phi=\half\gamma^{k\ell} A_kA_\ell -A_kU^k
\leqno(10)
$$
is the Newtonian potential relative to $V$. Conversely,
$A_b
=
-\gammaU_{bk}V^k+\Big(\phi-\half\gammaU_{k\ell} V^k V^\ell\Big)\theta_b$.

It must be emphasized that the `principle of general covariance'
has been be consistently adapted from general relativity to the NCB framework
[{\bf{DK}}], the specific Newtonian gauge group
$G=\Diff(M)\,\semidirectproduct\,(\Omega^1(M)\times\Fun(M))$ acting upon NCB
structures according to
$$
\pmatrix{
\gamma \ccr
\theta \ccr
U \ccr
V \ccr
\phi \ccr
}
\longmapsto
\cA_*\!
\pmatrix{
\gamma \hfill\ccr
\theta \hfill\ccr
U+\gamma(\Psi) \hfill\ccr
V+\gamma(d\cF) \hfill\ccr
\phi+V(\cF)+\half\gamma(d\cF,d\cF) \hfill\ccr
}
\leqno(11)
$$
where $\cA_*$ is the push-forward by $\cA\in\Diff(M)$, also
$\Psi\in\Omega^1(M)$ is a Galilei boost $1$-form and $\cF\in\Fun(M)$ a
special Newtonian gauge. As expected, $G$ does act on NC structures via
$\Diff(M)$ only
$$
\pmatrix{
\gamma \ccr
\theta \ccr
\Gamma \ccr
}
\longmapsto
\cA_*\!
\pmatrix{
\gamma \ccr
\theta \ccr
\Gamma \ccr
}.
\leqno(12)
$$

The infinitesimal action of the gauge group $G$ on a NCB structure thus
reads
$$
\delta\pmatrix{
\gamma \ccr
\theta \ccr
U \ccr
V \ccr
\phi \ccr
}
=\pmatrix{
L_\X\gamma \hfill\ccr
L_\X\theta \hfill\ccr
L_\X U+\gamma(\psi) \hfill\ccr
L_\X V+\gamma(df) \hfill\ccr
\X(\phi)+V(f) \hfill\ccr
}
\leqno(13)
$$
where $\X\in\Vect(M)$, $\psi\in\Omega^1(M)$ and
$f\in\Fun(M)$. The associated Lie algebra structure is explicitly given by
$$
\Big[(\X,\psi,f),(\X',\psi',f')\Big] =
\Big([\X,\X'],L_\X\psi'-L_{\X'}\psi,\X(f')-\X'(f)\Big).
\leqno(14)
$$

\goodbreak
\parag
i) {\sl The extended Coriolis algebra}

By looking at the infinitesimal NCB gauge transformations such that
$$
\delta\gamma=0,
\qquad
\delta\theta=0,
\qquad
\delta U=0
\leqno(15)
$$
we readily find a Lie algebra denoted by $\widetilde{\cor}(M,\gamma,\theta,U)$
that clearly consists of couples
$(\X,f)\in{\cor}(M,\gamma,\theta)\times\Fun(M)$
---the boosts are already fixed (modulo $\theta$): $\psi=\gammaU([U,\X])$.
The Lie brackets, inherited from Eq.~(14), reduce hence to
$$
\Big[(\X,f),(\X',f')\Big] =
\Big([\X,\X'],\X(f')-\X'(f)\Big)
\leqno(16)
$$
and yield the semidirect product structure
$$
\widetilde{\cor}(M,\gamma,\theta,U)
=
{\cor}(M,\gamma,\theta)\,\semidirectproduct\,\Fun(M).
\leqno(17)
$$

\goodbreak
\parag
ii) {\sl The extended Milne algebra}

Likewise, the Lie algebra $\widetilde{\mil}(M,\gamma,\theta,U,V)$
of all gauge transformations (13) such that
$$
\delta\gamma=0,
\qquad
\delta\theta=0,
\qquad
\delta U=0,
\qquad
\delta V=0
\leqno(18)
$$
is formed of pairs $(\X,\xi)$ with
$\X\in\mil(M,\gamma,\theta,\Gamma)$
and
$\xi\in\Fun(T)$
where $T\equiv{M/\ker(\theta)}$ is the canonical time axis ---the
general solution of $\gamma(df)=[V,\X]$ being of the form $f=\xi+f_\X$
with $f_\X$ uniquely determined by the condition $f_0=0$.
The Lie brackets
$$
\Big[
(\X,\xi),(\X',\xi')
\Big]
=
\Big(
[\X,\X'],\X(\xi'+f_{\X'})-\X'(\xi+f_{\X}) - f_{[\X,\X']}
\Big)
\leqno(19)
$$
therefore lead to the following {\it non central extension}
$$
0\to\Fun(T)\to\widetilde{\mil}(M,\gamma,\theta,U,V)\to
{\mil}(M,\gamma,\theta,\Gamma)\to 0.
\leqno(20)
$$

In the standard case and with the notation of section~1, the Lie brackets
$(\X'',\xi'')
=\Big[(\X,\xi),(\X',\xi')\Big]$
read
$$
\left\{
\eqalign{
\omega'' &= \omega'\omega-\omega\omega' \ccr
\varrho'' &= \omega'\varrho-\omega\varrho'
+\tau\dot{\varrho}'-\tau'\dot{\varrho} \ccr
\tau''&=0 \ccr
\xi'' &= \tau\dot{\xi}'-\tau'\dot{\xi}
+\varrho\cdot\dot\varrho'-\varrho'\cdot\dot\varrho \ccr
}
\right.
\leqno(21)
$$
with $\omega\in\so(n)$, $\varrho\in\Fun(T,\bR^n)$,
$\tau\in\bR$ and $\xi\in\Fun(T)$.

\goodbreak
\parag
iii) {\sl The extended Galilei algebra}

At last
$\widetilde{\gal}(M,\gamma,\theta,U,V,\phi)$
defined as the stabilizer
of a NCB structure, {\it viz}
$$
\delta\gamma=0,
\qquad
\delta\theta=0,
\qquad
\delta U=0,
\qquad
\delta V=0,
\qquad
\delta\phi=0
\leqno(22)
$$
consists of pairs $(\X,\xi)$
with $\X\in\gal(M,\gamma,\theta,\Gamma)$
and
$\xi\in\bR$ (the
overall additive constant in the solution $f$ of
$df
=
(-\X(\phi)+\gammaU(L_\X{V},V))\theta-\gammaU(L_\X{V})$).
The Lie brackets given by Eq.~(19) still hold and, this time, we find a non
trivial finite dimensional {\it central extension}
$$
0\to\bR\to\widetilde{\gal}(M,\gamma,\theta,U,V,\phi)\to
{\gal}(M,\gamma,\theta,\Gamma)\to 0.
\leqno(23)
$$

In the case of flat space\-time we get, with the same
notation as before, the centrally extended Galilei algebra
$$
\left\{
\eqalign{
\omega'' &= \omega'\omega-\omega\omega' \ccr
\beta'' &= \omega'\beta-\omega\beta' \ccr
\sigma'' &= \omega'\sigma-\omega\sigma'
+\beta'\tau-\beta\tau'\ccr
\tau''&=0 \ccr
\xi'' &= \sigma\cdot\beta'-\sigma'\cdot\beta \ccr
}
\right.
\leqno(24)
$$
i.e. the Lie algebra of the {\it Bargmann group}.

\bigskip

\parag
{\bf Acknowledgements}~:
Discussions with B.~Carter and J.-M.~Souriau are acknowledged. Thanks are due
to C.~Barrab\`es and P.~Horv\'athy for valuable help.

\vfill\eject



\centerline{\bf References}

\parag
[{\bf{CK}}]
B.~Carter and I.M.~Khalatnikov,
{\sl Canonically Covariant Formulation of Landau's Newtonian
Superfluid Dynamics}, Preprint DARC, Meudon (1992).

\parag
[{\bf{Ca}}]
E.~Cartan, 
{\sl Sur les vari\'et\'es \`a connexion affine et la
th\'eorie de la relativit\'e g\'en\'eralis\'e},
Ann. Sci. {\'E}cole Norm. Sup. (4)
{\bf 40}
(1923),
325--412.

\parag
[{\bf{K\"u}}]
H.P.~K{\"u}nzle, 
{\sl Galilei and {L}orentz structures on space\-time:
Comparison of the corresponding geometry and physics}, 
Ann. Inst. H. Poincar{\'e}.
Phys. Th{\'e}or.
{\bf 17}
(1972),  
337--362.

\parag
[{\bf{DK}}]
C.~Duval and H.P.~K\"unzle,
{\sl Sur les connexions newtoniennes et l'extension
non triviale du groupe de Galil\'ee},
C.R. Acad. Sci. Paris
{\bf 285 A}
(1977),
813--816;
{\sl Dynamics of continua and particles from
general covariance of Newtonian gravitation theory},
Rep. Math. Phys.
{\bf 13}
(1978),
351--368;
{\sl Minimal Gravitational Coupling in the Newtonian Theory
and the Covariant Schr\"odinger Equation},
Gen. Rel. Grav.
{\bf 16}
(1984),
333--347. 

\parag
[{\bf{Mi}}]
E.A.~Milne,
{\sl A Newtonian Expanding Universe},
Quat. J. Math. (Oxford Ser.)
{\bf 5}
(1934),
64--72.

\parag
[{\bf{Tra}}]
A.~Trautman,
{\sl Sur la th\'eorie newtonienne de la gravitation},
C.R. Acad. Sci. Paris
{\bf 257}
(1963),
617--620; 
{\sl Comparison of {N}ewtonian and relativistic theories of
space time}, pp.~413--425 in Perspectives in Geometry and
Relativity   (B.~Hoffmann, ed.),
Indiana University Press, Bloomington, 1964.

\parag
[{\bf{Tr\"u}}]
M.~Tr\"umper,
{\sl Lagrangian Mechanics and the Geometry of Configuration
Spacetime},
Ann. Phys. (N.Y.)
{\bf 149}
(1983),
203--233.

\bye